\documentclass[fleqn,10pt]{wlscirep}
\usepackage[utf8]{inputenc}
\usepackage[T1]{fontenc}
\title{Upper critical fields in normal metal--superconductor--normal metal trilayers}

\author[1]{Kelsey B. Robbins}
\author[1]{Pukar Sedai}
\author[1]{Alexandra J. Howzen}
\author[2]{Robert M. Klaes}
\author[2]{Reza Loloee}
\author[2]{Norman O. Birge}
\author[1,*]{Nathan Satchell}
\affil[1]{Department of Physics, Texas State University, San Marcos, Texas 78666, USA}
\affil[2]{Department of Physics and Astronomy, Michigan State University, East Lansing, Michigan 48824, USA}

\affil[*]{satchell@txstate.edu}

\begin{abstract}

The role of spin orbit interaction in superconducting proximity effect is an area of intense research effort. Recent theoretical and experimental works investigate the possible role of spin-orbit interaction in generating spin-triplet pair correlations. In this work, we present an experimental survey of thin normal metal--superconductor--normal metal trilayers with Nb superconductor and Al, Ti, Cu, Pt, Ta, and Au normal metals, along with single layers of Nb as reference. We aim to probe the role of spin-orbit interaction and resistivity on the normal metal proximity effect through measurements of the upper critical field. We find that the upper critical fields of the trilayers are lower than that of a single layer Nb reference sample, and that the trilayers with higher resistivity metals, Ti, Pt, and Ta, behave as 2-dimensional superconductors. At low applied in-plane magnetic fields and temperatures close to the zero field transition temperature, we find a possible deviation from 2-dimensional to 3-dimensional behavior in the Ti and Pt trilayers. We also find that compared to single layer Nb films, all of our trilayers show a greater suppression of critical temperature during rotation from an in-plane to an out-of-plane applied magnetic field, with the greatest suppression observed in trilayers with Au or Al. This suppression of the critical temperature under field rotation might appear analogous to the colossal spin valve effect that can be achieved in systems with ferromagnetic materials; however, in our trilayers, only conventional orbital screening contributions to the suppression are present and the additional suppression is not present in the absence of applied magnetic field.

\end{abstract}
\begin{document}

\flushbottom
\maketitle
 \thispagestyle{empty}

\section*{Introduction}

The superconducting proximity effect occurs when pair correlations from the superconductor proximitize adjacent materials and was first discovered in superconductor--normal metal--superconductor (S-N-S) pressed contacts by Holm and Meissner \cite{Holm1932}. For certain N layers, such as Cu, the proximity effect is long-ranged and can occur over several microns \cite{Meissner1960, clarke1968proximity}. Recently, the proximity effect in ferromagnetic materials has received attention due to the generation of spin-triplet correlations by the exchange energy, leading to the field of superconducting spintronics \cite{eschrig_review_2011, Eschrig_review_2015, linder_review_2015, melnikov_2022, cai_2023, birge_2024}. However, most observations of proximity effect in ferromagnetic materials are short-ranged in comparison to proximity in Cu.

Recent Josephson\cite{satchell_2018, Satchell_2019, Satchell_2020, Satchell_2021, komori_2021, kindiak_2024}, ferromagnetic resonance\cite{jeon_2018, Jeon_2019, Jeon_2020}, muon spin rotation\cite{flokstra_2023}, and electrical transport\cite{Banerjee_2018, Martinez_2020, Ruano_2020, bregazzi_2024} experiments have investigated superconductor-ferromagnet heterostructures with additional heavy normal metals, such as Pt, for the possible role of spin-orbit interaction in generating spin-triplets \cite{Niu_2012, Bergeret_2013, Bergeret_2014, Konschelle_2014, Jacobsen_2015, alidoust_2015, alidoust_2015b, jacobsen_2016, Costa_2017, Hikino_2018, Simensen_2018, Montiel_2018, Minutillo_2018, Johnsen_2019, Amundsen_2019, Eskilt_2019, Vezin_2020, Minutillo_2021, Olde_2021, Aunsmo_2024}, leading to a new field of superconducting spin-orbitronics. For an overview, see the review by Amundsen \textit{et al.} on spin-orbit effects in superconducting hybrid structures \cite{Amundsen2024}. 

Early theoretical work studied the upper critical field in S-N systems, assuming that the transparency of the interface was equal to unity \cite{Kupriyanov1985}. More recent theoretical work has shown that the upper critical field of N-S-N trilayers depends on the transparency of the interface and the normal coherence length for pair correlations \cite{Fominov_2001}. Previous experimental works studied the upper critical fields of Nb-Cu and Nb-Pd systems\cite{sidorenko_2002, cirillo_2003, kushnir_2006, kushnir_2009}. Layers with large spin-orbit interaction on the surface of thin superconducting layers have recently been predicted to provide a mechanism that can modify and tune the upper critical field \cite{Olde_2021}. Paschoa \textit{et al.} observed differences in the critical temperature of Nb/Pt bilayers compared to Nb/Cu bilayers under rotation from in-plane to out-of-plane applied magnetic field and interpreted their results as being due to the larger spin-orbit interaction in Pt relative to Cu\cite{paschoa_2024b}.

Motivated by observations of spin-triplet proximity effects in systems with heavy normal metals, we present an experimental survey of thin N-S-N trilayers. Nb is the superconducting layer and at 25~nm thickness, the superconductivity in Nb is strongly influenced by the proximity effect in adjacent N layers, which are fixed to a thickness of 15~nm. The chosen N layers are Al, Ti, Cu, Pt, Ta, and Au, to provide both light metals (weaker spin-orbit interaction) and heavy metals (stronger spin-orbit interaction). Our choice of materials not only allows us to examine the role of spin-orbit interaction but also allows us to explore the possible role of normal metal resistivity in the proximity effect. We measure the upper critical field of the N-S-N trilayers as a probe of the normal metal proximity effect and explore possible trends. 

Our measurements of the upper critical field may offer insight into the mechanism responsible for the breakdown of superconductivity in our trilayers under applied magnetic fields. Several mechanisms can drive this suppression, including conventional orbital screening, Zeeman splitting (Pauli limit), and, in some cases, additional effects from spin-orbit interaction or unconventional superconductivity, such as spin-triplet pairing. Orbital screening occurs when the applied field induces supercurrents that disrupt pair coherence, while the Pauli limit is reached when Zeeman splitting exceeds the superconducting gap, directly breaking spin-singlet pairs. The Pauli limit can be estimated following Clogston and Chandrasekhar as $B_P \approx 1.84\times T_c$ \cite{Clogston1962, Chandrasekhar1962}. In our samples, we show that orbital screening is the dominant mechanism of the upper critical field for both in-plane and out-of-plane applied fields.

\section*{Results}

\subsection*{Measurements of superconducting transition, $T_c$ and $H_{c2}$}

\begin{figure}[ht]
\centering
\includegraphics[width=\linewidth]{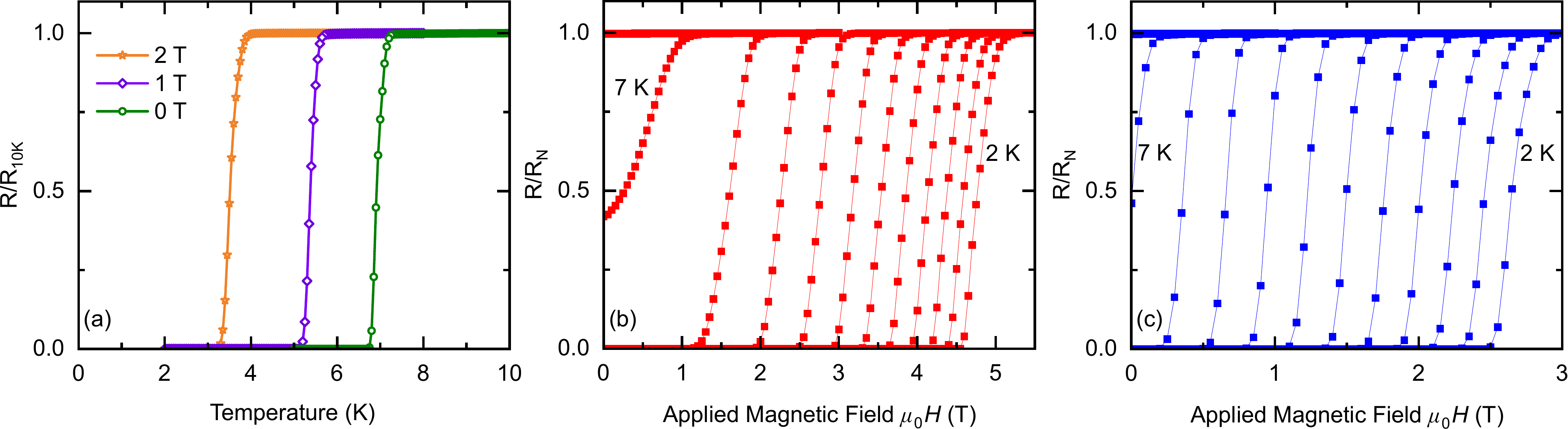}
\caption{Electrical measurements of a 25 nm single layer niobium film for (a) normalized resistance vs. temperature at 0, 1, and 2 T out-of-plane magnetic field; (b) normalized resistance vs. in-plane magnetic field; and (c) normalized resistance vs. out-of-plane magnetic field. Resistance vs. magnetic field measurements presented in (b) and (c) were taken at constant temperatures from 2 K in 0.5 K increments. Lines connecting the data are guides.}
\label{fig1}
\end{figure}

The superconducting transition is characterized using electrical transport measurements. Figure \ref{fig1} shows measurements on a single layer Nb (25~nm) sample, which we use as a reference and demonstration of how the data on all samples are acquired. Figure \ref{fig1} (a) shows the resistance as a function of temperature close to the transition for three different out-of-plane applied fields. Figure \ref{fig1} (b) and (c) show resistance as a function of applied magnetic field for (b) in-plane and (c) out-of-plane field orientations at fixed temperatures. We extract the critical parameter ($T_c$ or $H_{c2}$) at the 50\% normal state resistance value. There are two methods that we can use to extract the upper critical field of the superconductor with temperature, $H_{c2}$ ($T$). The first, using Figure \ref{fig1} (a), is to extract $H_{c2}$ ($T$) from measurements stepping temperature with a fixed magnetic field. Alternatively, we can fix the temperature of the sample and step magnetic field, shown in Figure \ref{fig1} (b) and (c). The data presented in Figure \ref{fig1} demonstrate that both methods provide a consistent value of $H_{c2}$ ($T$). From a practical point of view, we find that the second method has a significantly faster data acquisition time. We use both methods in this study and state which method was used in the relevant section.

\subsection*{The upper critical field, $H_{c2}$ ($T$)}

\begin{figure}[ht]
\centering
\includegraphics[width=\linewidth]{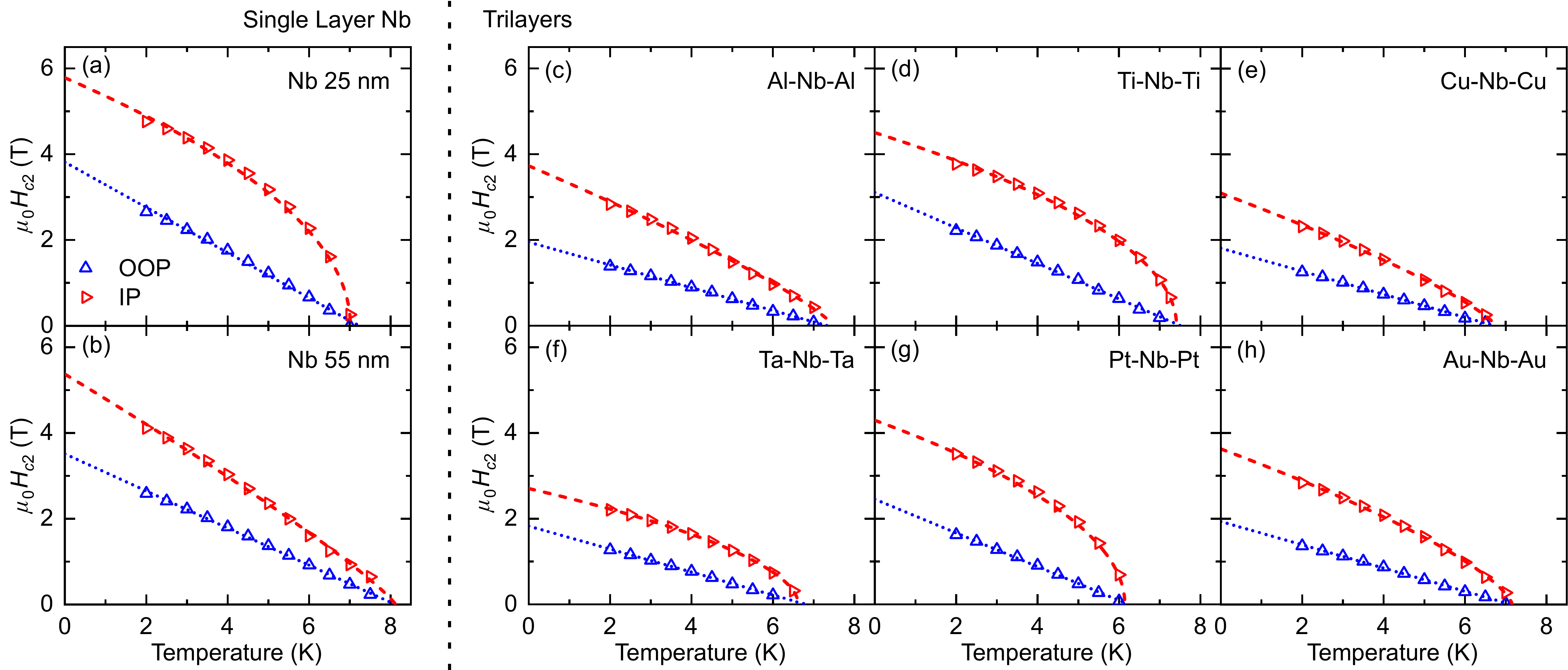}
\caption{Temperature dependance of the upper critical magnetic field, $H_{c2}$ for (a) 25 and (b) 55 nm thick, single layer Nb films and for (c) Al-Nb-Al, (d) Ti-Nb-Ti, (e) Cu-Nb-Cu, (f) Ta-Nb-Ta, (g) Pt-Nb-Pt, and (h) Au-Nb-Au trilayer films. The lines represent fits to the Ginzburg-Landau theory, Equation \ref{eq:Hc2T}. The best fit parameters are given in Table \ref{tabel}.}
\label{fig2}
\end{figure}

Figure~\ref{fig2} shows the extracted $H_{c2}$ ($T$) for the eight samples in this study determined by the method shown in Figure \ref{fig1} (b) and (c). We study two thicknesses of single layer Nb, 25~nm and 55~nm. The 25~nm thick Nb is chosen to match the thickness of the Nb in the N--S--N trilayers and 55~nm thick Nb is chosen to match the total thickness of the trilayer. We study six N--S--N trilayers where the N layers are Al, Ti, Cu, Pt, Ta, and Au, and are fixed at 15~nm thick on either side of the Nb. We follow the same measurement and analysis procedure for all samples in the study. The in-plane and out-of-plane applied field data are considered separately, and the temperature dependence of $H_{c2}$ ($T$) in both cases is fit to
\begin{equation}\label{eq:Hc2T}
{H_{c2} = H_{c2} (0~\text{K}) \left( 1-T/T_c (0~\text{T}) \right)^\alpha,}
\end{equation}

\noindent where $H_{c2} (0~\text{K})$ is the extrapolated zero temperature upper critical field and the exponent $\alpha$ is related to the film dimensionality. 

For out-of-plane applied fields, $\alpha$ is fixed as 1. The zero temperature isotropic coherence length $\xi_\text{GL}$ can be estimated from,
\begin{equation} \label{eq:coherence3D}
  H_{c2}^{\text{OOP}}(0~\text{K})  = \frac{\Phi_0}{2\pi \xi_\text{GL}^2},
\end{equation}

\noindent where $\Phi_0$ is the flux quantum. The best fit parameters and estimated $\xi_\text{GL}$ are given in Table~\ref{tabel}. For in-plane applied fields, the exponent $\alpha$ is related to dimensionality. $\alpha=1$ for 3D superconductors and $\alpha=0.5$ for 2D superconductors, and is a free fitting parameter between those bounds in our analysis. Again, the best fit parameters are given in Table~\ref{tabel}. 

For the 25~nm thick Nb sample, the best fit returns $\alpha=0.50\pm0.02$, consistent with 2D behavior. In the 2D limit of the Ginzburg-Landau model, $H_{c2}^{\text{IP}}(0~\text{K})$ can be related to the effective thickness of the superconductor, $d_\text{eff}$, via,
\begin{equation} \label{eq:coherence2D}
  H_{c2}^{\text{IP}}(0~\text{K})  = \frac{\sqrt{3} \Phi_0}{\pi d_\text{eff} \xi_\text{GL}}.
\end{equation}

\noindent Using our previously estimated $\xi_\text{GL}$, we estimate $d_\text{eff} = 21$~nm, reasonably consistent with the nominal 25~nm thickness of the Nb layer, particularly since it is known that Nb will form a native oxide and interface layers that decrease the layer's superconducting thickness \cite{Quarterman_2020}. For the 55~nm sample, $\alpha$ indicates a somewhat intermediate behavior, which is as expected as the thickness of the Nb film is increased. Upon increasing the thickness further, we would expect the Nb film to eventually show the 3D limit, $\alpha = 1$.

\begin{figure}[ht]
\centering
\includegraphics[width=\linewidth]{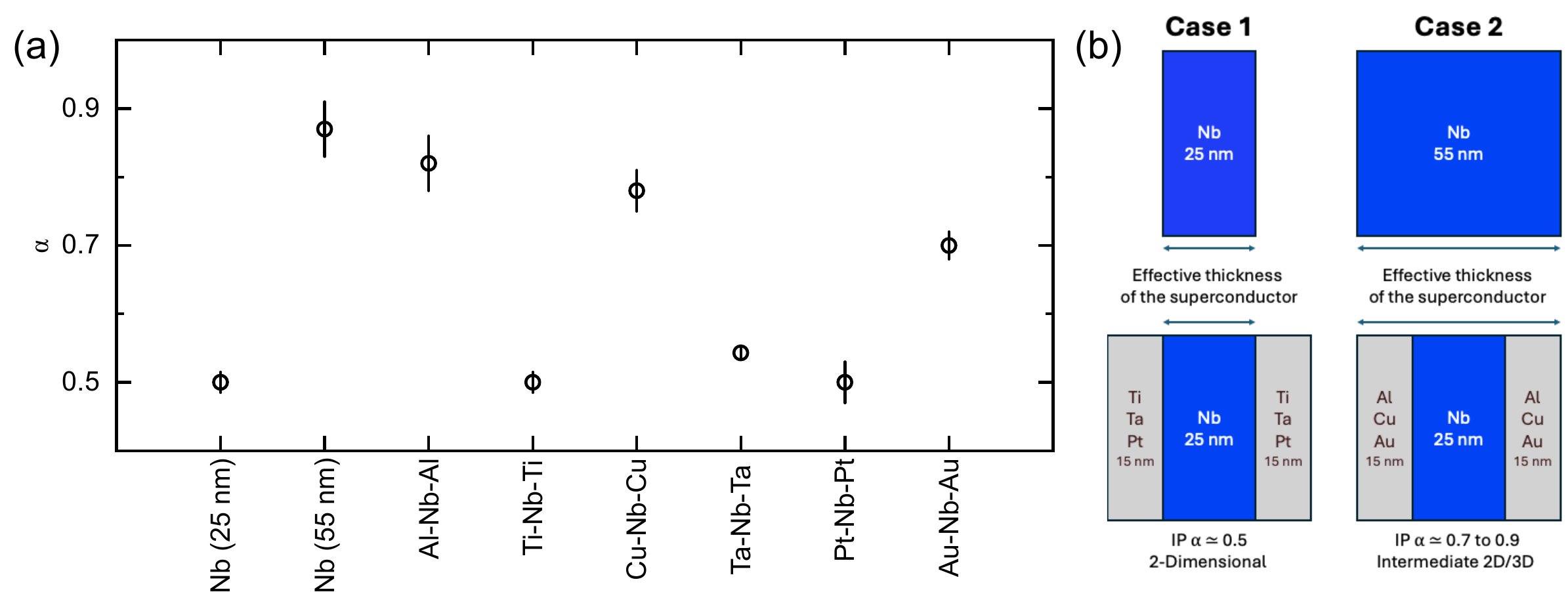}
\caption{Film dimensionality and effective thickness of the superconductor. (a) Extracted film dimensionality parameter, $\alpha$, for the fits presented in Figure \ref{fig2}. In the Ginzburg-Landau theory, $\alpha = 0.5$ indicates a 2-dimensional superconductor and $\alpha = 1$ a 3-dimensional superconductor. (b) Schematic representation of the effective thickness of the superconductor, $d_\text{eff}$, based on the comparison of fitted $\alpha$ and extracted $d_\text{eff}$ between the single layer and trilayer samples.}
\label{fig3}
\end{figure}

Having laid out the steps in our data collection and analysis, let us now turn to the N--S--N trilayers, shown in Figure~\ref{fig2} (c - h) and best fit parameters are again given in Table~\ref{tabel}. The proximity effect of normal metals clearly plays an important role in the modification of key parameters ($T_c$, $H_{c2}$, and $\alpha$), and these parameters are material dependent. We highlight the material dependence of $\alpha$ in Figure~\ref{fig3} (a).

\subsection*{Measurements of superconducting transition at low in-plane fields}

\begin{figure}[ht]
\centering
\includegraphics[width=0.75\linewidth]{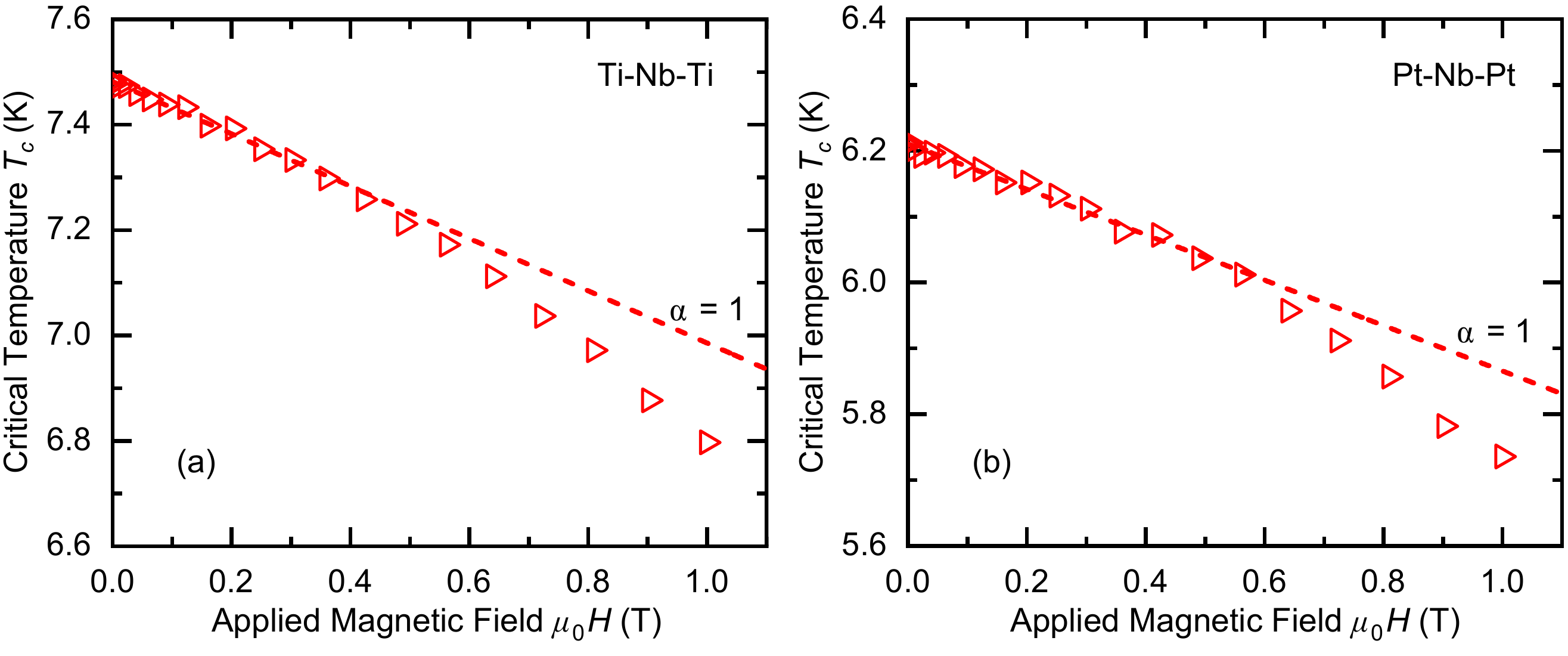}
\caption{Crossover from 3-dimensional towards 2-dimensional superconductivity at low applied in-plane magnetic fields in the (a) Ti-Nb-Ti and (b) Pt-Nb-Pt trilayers. The critical temperature is determined at each applied field value by measuring resistance while sweeping temperature at a fixed in-plane magnetic field. The dashed line represents Ginzburg-Landau theory, Equation~\ref{eq:TcH}, fit to the lowest field data only ($\mu_0H<0.4$~T) with $\alpha$ fixed to 1, the 3-dimensional limit.}
\label{fignew}
\end{figure}

We perform further measurements on the Ti and Pt trilayer samples in low in-plane applied magnetic fields to study behavior at $T$ close to $T_c$. We measure the $T_c$ while varying temperature, following the method shown in Figure \ref{fig1} (a), as this method allows for fine control over fixed low applied magnetic fields. Figure~\ref{fignew} shows the measured critical temperature of the Ti and Pt trilayers for in-plane applied fields. Equation \ref{eq:Hc2T} is rearranged to provide $T_c$ as a function of applied magnetic field,
\begin{equation}\label{eq:TcH}
{T_{c} (H) = T_{c} (0~\text{T}) \left( 1- \left( H/H_{c2} (0~\text{K}) \right)^{1/\alpha} \right),}
\end{equation}

\noindent where our new field-dependent parameter $T_c (H)$ corresponds to the original parameter $T$ in Equation~\ref{eq:Hc2T}, and $H$ is the value of the applied magnetic field.

\section*{Discussion}

The lateral dimensions of our films greatly exceed the penetration depth of polycrystalline Nb (about 100~nm\cite{Quarterman_2020}), but this length scale eclipses the total thickness of our films. For fields applied out-of-plane, the screening currents are in-plane and as a result the upper critical fields are well described by fixing the dimensionality parameter $\alpha$, in Equation \ref{eq:Hc2T}, to 1. For fields applied in-plane, the screening currents are out-of-plane and are strongly influenced by the interaction of the supercurrent with the surfaces and the impedance provided by the S/N interfaces. Based on $\alpha$, it is possible to group the behavior of the single Nb layers and the N-S-N trilayers into two categories. The 25~nm Nb layer and the Ti, Ta, and Pt trilayers are in the 2D limit with $\alpha \approx 0.5$. The 55~nm Nb layer and the Al, Cu and Au trilayers are in an intermediate regime with $0.7 \leq \alpha \leq 0.9$. None of the samples in this study fall into the 3D limit; however, previous work suggests that much thicker 200~nm Nb films can be described by $\alpha = 1$ \cite{Quarterman_2020}.

The dependence of $\alpha$ appears to correlate with the resistivity of the normal metal in the trilayer. To highlight this trend, we provide the bulk resistivity values for the normal metals in Table~\ref{tabel}. The higher resistivity metals, Ti, Ta, and Pt, correspond to the trilayers showing 2D behavior. The 2D behavior of the Ti, Ta, and Pt trilayers suggests that the screening currents are more confined to the Nb layer, and do not spread as much into the normal metals. On the other hand, for the Al, Cu, and Au trilayers, the screening currents occupy the whole thickness of the trilayer. Shown schematically in Figure \ref{fig3} (b), the trilayers can be thought of as having an effective superconducting thickness that is either similar to the thickness of only the Nb layer (Case 1) or similar to the thickness of the entire trilayer (Case 2). Extracting $d_\text{eff}$ from Equation~\ref{eq:coherence2D} yields 25, 31, and 23~nm for the Ti, Ta, and Pt trilayers, respectively, which is consistent with the interpretation that screening currents are more confined in these trilayers. It is possible that the higher resistivity of Ti, Ta, and Pt suppresses the screening currents in these layers.

The other extracted superconducting parameters, $H_{c2}$ and $T_c$, do not follow the same trend or grouping by material as $\alpha$, and do not appear to correlate with the resistivity of the normal metal or the strength of the spin-orbit interaction. For the critical temperature of the trilayer, compared to the single layer Nb film, we expect there to be some competition between an increase in $T_c$ due to normal metals acting as a capping layer to prevent the oxidation of Nb and the influence due to the proximity effect\cite{Fominov_2001} or spin-obit interaction\cite{Ptok_2018}. Numerous Nb alloys are known to exhibit intrinsic superconductivity. Given the elevated $T_c$ observed in our Ti and Al trilayers, it is possible that the formation of alloy layers at the interfaces with these normal metals contributed to the increase in $T_c$.

For $T$ close to $T_c$, the expected behavior of thin superconducting films is for the superconductivity to be 2D, as $\xi$ diverges at high temperatures and exceeds the thickness of the layer \cite{Schneider1991}. To study whether this expectation holds for the trilayers, we undertake further low field measurements on the Ti and Pt trilayer samples shown in Figure~\ref{fignew}. Our experimental finding is that for $T$ close to $T_c$, the Ti and Pt trilayers are best described by fixing $\alpha$ to the 3D limit of 1, deviating at higher fields towards 2D behavior. The exact mechanism for this behavior is not known. We note that deviation of $H_{c2}$ at $T$ close to $T_c$ has been previously reported in both thin film and bulk Nb\cite{PhysRevB.93.184501, PhysRevB.44.7585}.

For the upper critical field, although strong spin-orbit interaction has been shown to increase the in-plane critical field of ultra-thin 2D superconductors\cite{PhysRevB.25.171, Nam_2016}, we do not see any evidence of such an enhancement of $H_c$ here. Indeed, for all N-S-N trilayers, the upper critical fields are lower than that of the single layer Nb films we study as reference. Over the range of temperatures we study, we do not observe any signatures suggesting modification or tuning of the upper critical field by spin-orbit interaction\cite{Olde_2021}, as the data are well described by Equation~\ref{eq:Hc2T}. Revisiting the mechanisms responsible for the breakdown of superconductivity: for our 25~nm-thick Nb, the estimated Pauli limit is 13~T, which is much higher than the largest critical fields measured in our samples. This suggests that, within the applied field range of this study, superconductivity is primarily suppressed by orbital screening. There also does not appear to be a correlation between $T_c$ and $H_{c2}$. For example, the Pt trilayer has a $T_c$ that is significantly lower than the other trilayers, but the second highest $H_{c2}$ behind only the Ti trilayer, which has the highest $T_c$.

In superconducting systems containing ferromagnetic materials, the spin valve effect is the suppression or enhancement of the critical temperature ($T_c$) by rotating two (or more) ferromagnetic layers with respect to each other from a parallel to antiparallel or perpendicular alignment \cite{oh_1997, Tagirov_1999, buzdin_1999, fominov_2010, Mironov_2014}. In the perpendicular alignment, the suppression of $T_c$ is due to the generation of spin-triplets. The spin valve effect manifests at zero applied magnetic field, provided that the anisotropy of the ferromagnets is sufficient to maintain alignment at remanence. One experimental realization is performed by rotating the sample from field applied in-plane to out-of-plane to achieve a perpendicular magnetic alignment \cite{Singh_2015, Voltan_2016, feng_2017}. In the previous work of Singh \textit{et al.}, the colossal spin valve effect is achieved when the Ni layer in their structure is rotated by an applied out-of-plane magnetic field to be perpendicular to the CrO$_2$ layer, generating spin-triplets. It is generally the case in thin-films that the geometry of the layers causes a lowering of $T_c$ for out-of-plane fields compared to in-plane fields because of the increased orbital screening suppression when the field is out-of-plane. The orbital screening suppression can be removed to isolate spin-triplet contributions by measuring at zero applied field, however it is not always possible to maintain alignment of the ferromagnetic layers without a field, due to magnetic anisotropy. Therefore, Singh \textit{et al.} compared their spin valve to reference samples, such as single-layer S films, to isolate the spin-triplet component of their effect \cite{Singh_2015}. Paschoa \textit{et al.} interpret their data on S--N bilayers in analogy to the colossal spin valve effect\cite{paschoa_2024b}.

\begin{figure}[ht]
\centering
\includegraphics[width=\linewidth]{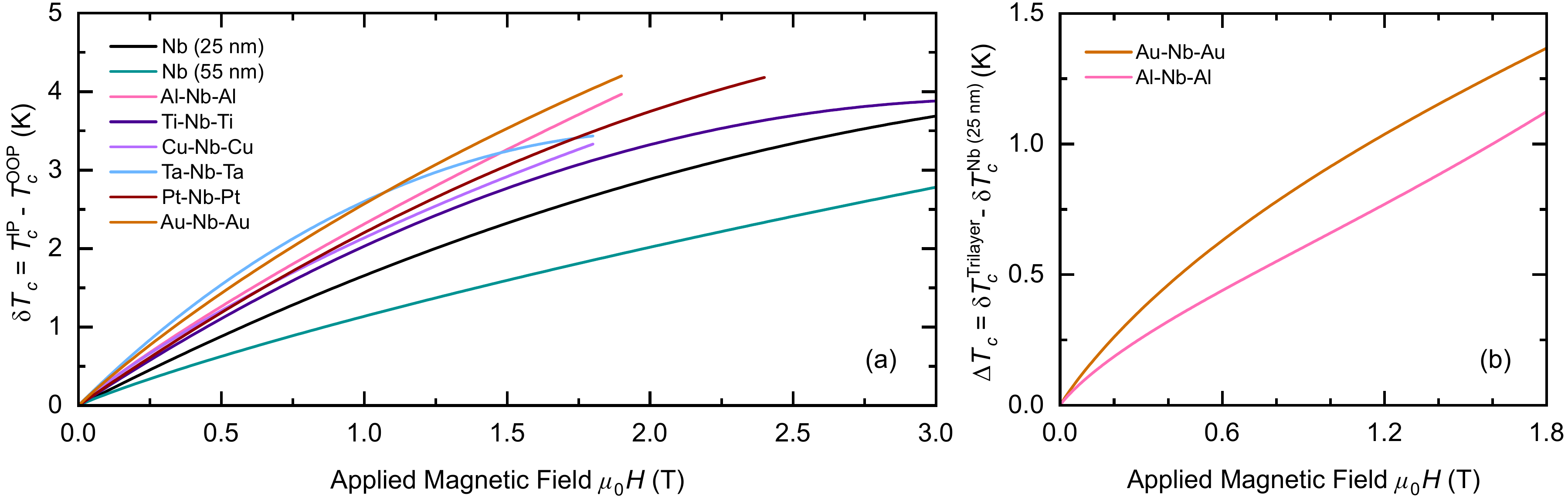}
\caption{Ostensible analogy with the colossal spin-valve effect. (a) The extracted change in critical temperature, $\delta T_c$, between in-plane and out-of-plane applied magnetic fields. The trilayers show a greater suppression of $T_c$ compared to the single layer Nb films. (b) The difference in $\delta T_c$ between the trilayers and 25~nm single layer Nb film, $\Delta T_c$, for the Au and Al trilayers.}
\label{fig4}
\end{figure}

Because our samples undergo a similar rotation from in-plane to out-of-plane applied fields as those of Singh \textit{et al.}, we discuss the ostensible analogy between the colossal spin valve effect and the orbital screening suppression in our trilayers. Equation \ref{eq:TcH} provides analytically $T_c$ as a function of applied magnetic field. For this analysis, we consider only the high field data from Figure \ref{fig2} and Table~\ref{tabel}, as this is where the changes in $T_c$ with applied field are most pronounced. Next, we define a change of $T_c$ between in-plane and out-of-plane applied fields as, $ \delta T_{c}  = T_{c}^{\text{IP}} - T_{c}^{\text{OOP}}.$ $\delta T_{c}$ is plotted for all samples in this study in Figure \ref{fig4} (a). We report that $\delta T_{c}$ is greater for all trilayer films compared to single-layer Nb films, demonstrating that normal metals contribute an increase to the orbital screening suppression of $T_c$. The largest observed $\delta T_{c}$ in the N-S-N trilayers is achieved for Au or Ta (depending on the applied field value) and smallest $\delta T_{c}$ for Ti. At high fields, the Al trilayer shows the second largest suppression in $T_c$, highlighting that there is no trend in $\delta T_{c}$ with the strength of spin-orbit interaction. To continue our analogy by comparing to a reference sample, we plot in Figure~\ref{fig4} (b) $\Delta T_c$, the difference between the trilayers with the single layer Nb for Au and Al. Doing so isolates only the contribution of additional suppression of $T_c$ due to the presence of either Au or Al layers. Although our purely orbital screening effect is not a spin valve effect (it cannot manifest at zero applied field), it is interesting to note that the additional suppression due to the Au layers, $\Delta T_c \approx 1.3$~K at 1.8~T, is greater than the additional suppression due to spin-triplets, $\Delta T_c \approx 0.8$~K at 2.0~T, observed by Singh \textit{et al.}\cite{Singh_2015}. In our trilayers, $\delta T_{c}$ is clearly material dependent; however, there does not appear to be a clear trend indicating a role for normal metal resistivity or spin-orbit interaction, contradicting the conclusions of the recent work of Paschoa \textit{et al.} on Nb/Pt bilayers\cite{paschoa_2024b}

\section*{Conclusions}

In conclusion, we study a series of N-S-N trilayers with Al, Ti, Cu, Ta, Pt, and Au normal metals, along with single layers of Nb as reference, to probe the role of the proximity effect on the upper critical field of the trilayers. The upper critical fields of all our samples are well described by the standard Ginzberg-Landau functions, and we extract key parameters and trends from this framework. Trilayers with Ti, Ta, and Pt as well as a 25~nm thick Nb layer show responses to in-plane applied magnetic fields that are consistent with what would be expected for a 2-dimensional superconductor. We speculate that it is the higher resistivity of these normal metals, compared to Al, Cu, and Au, that causes a confinement of the superconductivity and hence the 2-dimensional behavior. We do not find strong trends in the critical temperatures or upper critical fields of the trilayers. The upper critical fields of the trilayers are suppressed relative to those of the Nb reference samples. At low applied magnetic fields and temperatures close to the zero field transition temperature, we found a possible deviation from 2-dimensional to 3-dimensional behavior in the Ti and Pt trilayers. We calculate the change in critical temperature of the trilayers under a rotation from in-plane to out-of-plane applied magnetic field and find that the trilayers have enhanced suppression of $T_c$ during rotation relative to single layers of Nb. We discuss an ostensible analogy with the colossal spin-valve effect.
\section*{Methods}

The films were deposited in a single vacuum cycle using dc sputtering in a vacuum system with a base pressure of $2 \times 10^{-8}$~Torr and a partial water pressure of $3 \times 10^{-9}$~Torr after liquid nitrogen cooling. The samples were grown on 0.4 mm thick Si substrates with 100~nm thermal oxide. The growth was carried out at an approximate Ar pressure of 2~mTorr, at a typical growth rate of 0.4 nm s$^{-1}$. Growth rates were calibrated using an \textit{in situ} crystal film thickness monitor to provide nominal layer thicknesses.

Resistivity measurements were performed in a Quantum Design Dynacool Physical Property Measurement System, equipped with a 9~T perpendicular field, and the Horizontal Rotator and Resistivity options. 4-point-probe electrical connections were made in a Van der Pauw geometry using either indium soldering or pressed contacts. For all measurements, a current of 1~mA was applied. The sample was rotated at a rate of 1$^\circ$/s between in-plane and out-of-plane applied magnetic fields \textit{in situ}.

\section*{Data Availability}

The datasets generated during and/or analyzed during the current study are available in the Texas State University Dataverse Repository, https://doi.org/10.18738/T8/0MGNEY.

\bibliography{sample}

\section*{Acknowledgments}

We acknowledge experimental assistance through the Analysis Research Service Center from Sam Cantrell and Casey Smith. We acknowledge support from new faculty startup funding made available by Texas State University.

\section*{Author contributions statement}

N.S. conceived the study and wrote the manuscript. K.B.R. conducted the electrical transport measurements. K.B.R, A.J.H. and N.S. analyzed the data. P.S., R.M.K., R.L., N.O.B. and N.S. undertook the materials development and sample growth. All authors reviewed the manuscript. 

\section*{Additional information}

The authors declare no competing interests.

\newpage

\begin{table}[]
\resizebox{\columnwidth}{!}{%
\begin{tabular}{|l|c|c|c|c|c|c|c|c|}
\hline
Sample &
  \begin{tabular}[c]{@{}c@{}}OOP $T_c$ \\ (K)\end{tabular} &
  \begin{tabular}[c]{@{}c@{}}OOP $\mu_0H_{c2}$ \\ (T)\end{tabular} &
  \begin{tabular}[c]{@{}c@{}}OOP $\alpha$ \\ (fixed)\end{tabular} &
  \begin{tabular}[c]{@{}c@{}}IP $T_c$ \\ (K)\end{tabular} &
  \begin{tabular}[c]{@{}c@{}}IP $\mu_0H_{c2}$ \\ (T)\end{tabular} &
  \begin{tabular}[c]{@{}c@{}}IP $\alpha$\\ ($0.5\leq \alpha \leq 1$)\end{tabular} &
  \begin{tabular}[c]{@{}c@{}}$\xi_\text{GL}$ \\ (nm)\end{tabular} &
  \begin{tabular}[c]{@{}c@{}}Bulk Resistivity \\ ($\mu$Ohm cm)\end{tabular} \\ \hline
Nb (25 nm) & 7.23 $\pm$ 0.08 & 3.81 $\pm$ 0.06 & 1 & 7.02 $\pm$ 0.02  & 5.78 $\pm$ 0.08 & 0.50 $\pm$ 0.02    & 9.29 &      \\ \hline
Nb (55 nm) & 8.11 $\pm$ 0.04 & 3.51 $\pm$ 0.03 & 1 & 8.1 $\pm$ 0.1    & 5.37 $\pm$ 0.09 & 0.87 $\pm$ 0.05   & 9.68 &      \\ \hline
Al-Nb-Al   & 7.34 $\pm$ 0.04 & 1.95 $\pm$ 0.02 & 1 & 7.5 $\pm$ 0.1    & 3.73 $\pm$ 0.07 & 0.82 $\pm$ 0.04   & 13.0 & 2.74 \\ \hline
Ti-Nb-Ti   & 7.51 $\pm$ 0.05 & 3.11 $\pm$ 0.03 & 1 & 7.4 $\pm$ 0.2    & 4.5 $\pm$ 0.1   & 0.50 $\pm$ 0.05   & 10.3 & 43.1 \\ \hline
Cu-Nb-Cu   & 6.73 $\pm$ 0.03 & 1.81 $\pm$ 0.01 & 1 & 6.75 $\pm$ 0.05  & 3.09 $\pm$ 0.04 & 0.78 $\pm$ 0.03   & 13.5 & 1.7  \\ \hline
Ta-Nb-Ta   & 6.82 $\pm$ 0.04 & 1.83 $\pm$ 0.01 & 1 & 6.63 $\pm$ 0.01  & 2.70 $\pm$ 0.01  & 0.543 $\pm$ 0.007 & 13.4 & 13.1 \\ \hline
Pt-Nb-Pt   & 6.24 $\pm$ 0.04 & 2.45 $\pm$ 0.03 & 1 & 6.14  $\pm$ 0.05 & 4.29 $\pm$ 0.09 & 0.50 $\pm$ 0.03   & 11.6 & 10.4 \\ \hline
Au-Nb-Au   & 7.13 $\pm$ 0.04 & 1.93 $\pm$ 0.02 & 1 & 7.16 $\pm$ 0.04  & 3.63 $\pm$ 0.04 & 0.70 $\pm$ 0.02   & 13.1 & 2.2  \\ \hline
\end{tabular}%
}
\caption{\label{tabel}Best fit parameters of Equation~\ref{eq:Hc2T} to the data shown in Figure~\ref{fig2}. The parameters are grouped depending on whether they correspond to fitting the in-plane (IP) or out-of-plane (OOP) applied magnetic field data. The Ginzberg-Landau coherence length, $\xi_\text{GL}$, is estimated from Equation~\ref{eq:coherence3D}. Bulk resistivity values are reproduced from Kittel\cite{Kittel}.}
\end{table}

\end{document}